\documentclass{aastex6}
\usepackage{multirow}
\usepackage{color}

\newcommand{\kms}{\rm \,km\,s^{-1}}

\newcommand{\msun}{M_\odot}

\fullcollaborationName{The Friends of AASTeX Collaboration}

\begin{document}




\title{Triggered O star formation in M\,20 via cloud-cloud collision: \\ Comparisons between high-resolution CO observations and simulations}

\author{K. Torii\altaffilmark{1}, Y. Hattori\altaffilmark{2}, K. Hasegawa\altaffilmark{2}, A. Ohama\altaffilmark{2}, T. J. Haworth\altaffilmark{3}, K. Shima\altaffilmark{4}, A. Habe\altaffilmark{4}, K. Tachihara\altaffilmark{2}, N. Mizuno\altaffilmark{5},  T. Onishi\altaffilmark{6}, A. Mizuno\altaffilmark{7}, and Y. Fukui\altaffilmark{2}}
\affil{$^1$Nobeyama Radio Observatory, 462-2 Nobeyama Minamimaki-mura, Minamisaku-gun, Nagano 384-1305, Japan}
\affil{$^2$Department of Physics, Nagoya University, Chikusa-ku, Nagoya, Aichi 464-8601, Japan}
\affil{$^3$Astrophysics Group, Imperial College London, Blackett Laboratory, Prince Consort Road, London SW7 2AZ, UK}
\affil{$^4$Faculty of Science, Department of Physics, Hokkaido University, Kita 10 Nishi 8 Kita-ku, Sapporo 060-0810, Japan}
\affil{$^5$National Astronomical Observatory of Japan, Mitaka, Tokyo 181-8588, Japan}
\affil{$^6$Department of Astrophysics, Graduate School of Science, Osaka Prefecture University, 1-1 Gakuen-cho, Nakaku, Sakai, Osaka 599-8531, Japan}
\affil{$^7$Solar-Terrestrial Environment Laboratory, Nagoya University, Chikusa-ku, Nagoya 464-8601, Japan}


\begin{abstract}
High-mass star formation is one of the top-priority issues in astrophysics. Recent observational studies are revealing that cloud-cloud collisions may play a role in high-mass star formation in several places in the Milky Way and the Large Magellanic Cloud. 
The Trifid Nebula M\,20 is a well known galactic H{\sc ii} region ionized by a single O7.5 star. 
In 2011, based on the CO observations with NANTEN2 we reported that the O star was formed by the collision between two molecular clouds $\sim$0.3\,Myr ago. 
Those observations identified two molecular clouds towards M\,20, traveling at a relative velocity of $7.5\kms$. 
This velocity separation implies that the clouds cannot be gravitationally bound to M20, but since the clouds show signs of heating by the stars there they must be spatially coincident with it. A collision is therefore highly possible. 
In this paper we present the new CO $J$=1--0 and $J$=3--2 observations of the colliding clouds in M\,20 performed with the Mopra and ASTE telescopes.
The high resolution observations revealed the two molecular clouds have peculiar spatial and velocity structures, i.e., the spatially complementary distribution between the two clouds and the bridge feature which connects the two clouds in velocity space.
Based on a new comparison with numerical models, we find that this complementary distribution is an expected outcome of cloud-cloud collisions, and that the bridge feature can be interpreted as the turbulent gas excited at the interface of the collision.
Our results reinforce the cloud-cloud collision scenario in M\,20.

\end{abstract}





\section{Introduction} \label{sec:intro}

There is increasing evidence that cloud-cloud collision plays an important role in the high-mass star formation. 
Based on CO observations with the NANTEN2 4-m telescope, \citet{fur2009} and \citet{oha2010} revealed that two giant molecular clouds having a velocity separation of 20\,$\kms$ are both associated with the H{\sc ii} region RCW\,49, which is excited by the massive star cluster Westerlund\,2.
The large velocity separation cannot be interpreted either as the gravitational binding or as the expanding motion driven by the stellar feedback. 
An alternative proposed by the authors is a scenario in which the massive cluster was formed by a supersonic collision between the two clouds, where the observed velocity separation can be deemed as the projection of the colliding velocity.
Following the discovery in Westerlund\,2, the association between two clouds of significantly different velocities with O stars were reported in several high-mass star forming regions in the Milky Way, and triggered O star formation by cloud-cloud collision was discussed as a plausible interpretation; e.g., \citet[][hereafter paper I]{tor2011} and \citet{tor2015} for the galactic H{\sc ii} regions M\,20 and RCW\,120, and \citet{fuk2014} and \citet{fuk2016a} for the massive star clusters NGC\,3603 and RCW\,38.
In the Large Magellanic Cloud, ALMA observations led to the discoveries of a 36\,$\msun$ star in N159 West and a 40\,$\msun$ star in N159 East, likely triggered by collisions between filamentary clouds \citep{fuk2015b, sai2016}. 
Most recently, Fukui et al. (2016b in prep.) proposed a scenario that the Orion Nebula Cluster (ONC) was formed by a collision between two molecular clouds.
These results suggest that cloud-cloud collisions can trigger the formation of O stars for a wide mass range. 
M\,20 and RCW\,120 are H{\sc ii} regions dominated by a single O star, while Westerlund\,2, NGC\,3603, RCW\,38, and the ONC are massive star clusters which harbor several or more than ten O stars. 
\citet{fuk2016a} discussed that H$_2$ column density of the clouds is a critical parameter to determine the difference and that 10$^{23}$\,cm$^{-2}$ is required to form a massive star cluster.

\begin{figure}
\epsscale{.85}
\plotone{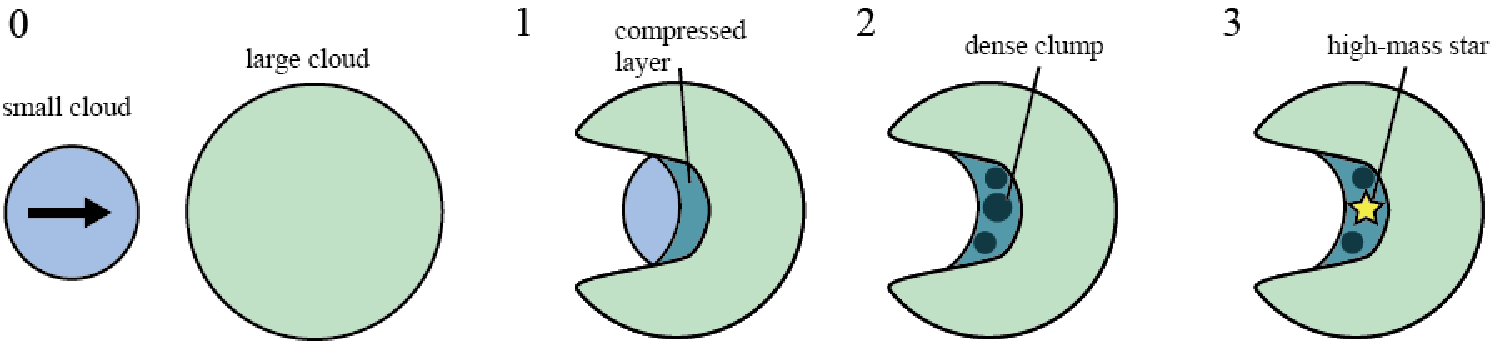}
\caption{Schematics of the cloud-cloud collision between two dissimilar clouds simulated by \citet{hab1992}.
 \label{sch}}
\end{figure}

A pioneering study of the numerical calculations of cloud-cloud collision was performed by \citet{hab1992}, followed by \citet{ana2010} and \citet{tak2014}. 
These authors simulated head-on collisions between two molecular clouds with different sizes.
Figure\,\ref{sch} shows a schematic picture of high-mass star formation resulting from a cloud-cloud collision between two dissimilar clouds, although the connection between cloud-cloud collision and high-mass star formation was not discussed in depth in the aforementioned simulations.
These authors indicate that cloud-cloud collision can induce the formation of the dense self-gravitating clumps inside the dense gas layer formed at the interface of the collision.
Formation of the massive clumps in the collisional-compressed layer was also discussed in depth in the recent magneto-hydrodynamical (MHD) simulations by \citet{ino2013}.
In their simulations, supersonic collisions amplify the magnetic field and the turbulent velocity of the gas, leading to a mass accretion rate of $10^{-4}$\,--\,$10^{-3}\,\msun$\,yr$^{-1}$, two orders of magnitude higher than that in the case of low-mass star formation.
Such a high mass accretion rate satisfies the conditions for high mass star formation as postulated in theoretical works \citep[e.g.,][]{mck2003, hos2010}.
One interesting signature characteristic to the collision between two dissimilar clouds is the cavity created on the surface of the large cloud.
The size of the cavity corresponds to the diameter of the small cloud, and the depth of the cavity is determined by the timescale of the collision and the balance of the momenta between the two colliding clouds.

An observational support for the cavity creation and the subsequent O star formation predicted in the cloud-cloud collision model was first given by \citet{tor2015} in the H{\sc ii} region RCW\,120.
The authors identified two molecular clouds with a velocity separation of $\sim20\kms$ and discussed that the exciting O star inside the RCW\,120's bright mid-infrared ring was formed through the collision of the two clouds.
The RCW\,120's beautiful ring is usually discussed to be formed by the expansion of the H{\sc ii} region. \citet{tor2015} however found no evidence of the expanding motion and discussed that the observed ring emission can be interpreted as the cavity created through the cloud-cloud collision.

Another observational signature of cloud-cloud collision is the ``bridge feature'' seen in a position-velocity diagram. 
Using the model data of the cloud-cloud collisions calculated by \citet{tak2014}, \citet{haw2015b, haw2015} conducted synthetic CO line observations and found a broad intermediate velocity feature which bridges between two colliding clouds in the position-velocity diagram.
The bridge feature probes the turbulent motion of the gas enhanced by the collision.
The bridge feature was observationally confirmed in the young massive star cluster RCW\,38 by \citet{fuk2016a}.
It was identified at a spot very nearby the O stars in RCW\,38, suggesting that the cloud-cloud collision in RCW\,38 is still continuing.

\begin{figure}
\epsscale{.85}
\plotone{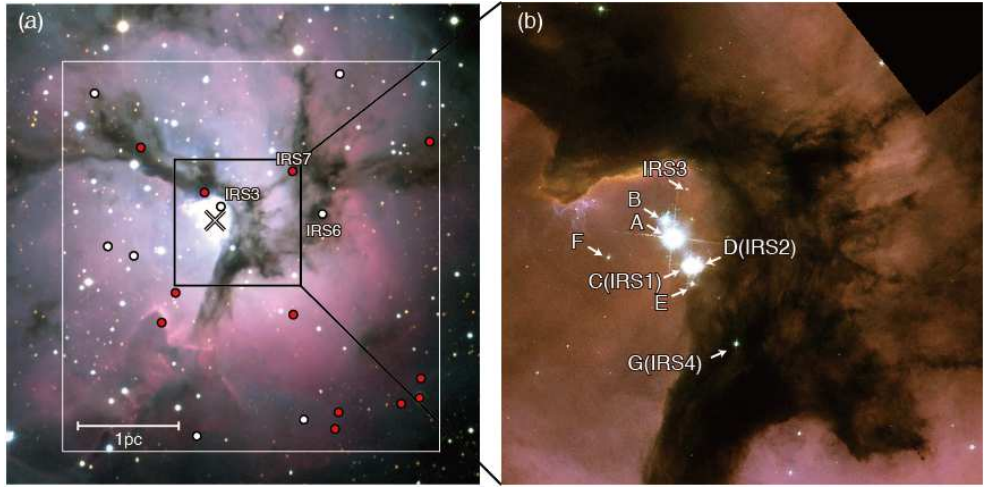}
\caption{(a) Optical image of M\,20, trisected by three dark dust lanes. (credit: NOAO). The exciting O7 star (HD\,164492\,A) is depicted by cross, while Class I/0 and Class II young stars identified by the {\it Spitzer} color-color diagram \citep{rho2006} are plotted with filled red circles and filled white circles, respectively. White solid lines show the target region of the present work. (b) The HST image of the central region of M\,20 \citep{yus2005a}. The HD\,164492 components listed in Table\,{tab:1} are indicated by arrows.
 \label{opt}}
\end{figure}

Among the previously studied cloud-cloud collision regions, M\,20, also known as the Trifid Nebula, is the youngest object along with RCW\,38 \citep[see review by][]{rho2008}. 
It was formed only 0.3\,Myr ago \citep{cer1998}. 
M\,20 has an outstanding obscuring dust lanes which trisect a nebula of gas ionized by an O7 star (HD\,164492\,A), which dominates the excitation of M\,20 (Figure\,\ref{opt}). 
The distance to M\,20 estimated in the previous studies ranges from 1.7\,kpc in the Sagittarius arm \citep{lyn1985} to 2.7\,kpc in the Scutum arm \citep{cam2011}.
In this paper we tentatively assume the distance of 1.7\,kpc to make our analysis and discussion consistent.
As shown in Figure\,\ref{opt}(b) and Talbe\,\ref{tab:1}, HD\,164492\,A is accompanied by several early type stars, forming a small stellar group within the central $\sim0.1$\,pc \citep{koh1999}.
The total stellar mass within $\sim$3\,pc of HD\,164492\,A is about 500\,$M_\odot$ \citep{ogu1975, rho2001, bro2013}, two orders of magnitude less massive than the massive star clusters like Westerlund\,2. 

Low-mass stars at different stages of star formation have been detected throughout the nebula at various wavelengths; i.e., optical jets \citep{cer1998, hes2004}, mid- and far-infrared young stellar objects (YSOs) \citep{rho2006}, infrared and X-ray YSOs \citep{rho2001, rho2004, lef2001, fei2013}, H$\alpha$ emission stars \citep{her1957, yus2005a}.
\citet{rho2001} identified 85 T-Tauri stars in the nebula. 
\citet{rho2006} cataloged $\sim160$ YSOs based on the {\it Spitzer} observations and classified them into different evolutionary stages.
Recently, the MYStIX project utilized the near-infrared and X-ray observations to identify more than 500 YSOs in M\,20 \citep{fei2013, bro2013, kuh2013}.
In Figure\,\ref{opt}(a) the class 0/I YSOs (red circles) and the class II YSOs (white circles) identified by \citet{rho2006} are plotted on the optical image of M\,20.
\citet{lef2008} made comparisons between the \citet{rho2006}'s YSOs and sub-mm dust continuum emission, indicating that many of the class 0/I YSOs are embedded within the dense molecular clumps.
On the other hand, the ISOCAM observations by \citet{lef2001} unveiled four point-like infrared sources (dubbed IRS\,2, IRS\,3, IRS\,6, and IRS\,7) with bright emission in the 9.7\,$\mu$m silicate band, direct evidence of the evaporating disk phase ``proplyd'', a later evolutionary phase of the star formation (Table\,\ref{tab:1}).
IRS\,3, IRS\,6, and IRS\,7 are also identified in the \citet{rho2006}'s YSO catalog (see Figure\,\ref{opt}(a)).

Thanks to the youth of M\,20, the natal molecular clouds have been less dissipated by the UV radiation, providing a unique opportunity to investigate the formation mechanism of the O star.
In Paper I we made CO $J$=1--0 and $J$=2--1 observations with NANTEN and NANTEN2 toward a large area of M\,20 at spatial resolutions of 90$''$ and 156$''$, revealing that two molecular clouds with $\sim10^3$\,$M_\odot$ having velocity separation of $\sim7\kms$ are both physically associated with the H{\sc ii} region in M\,20.
The two clouds both have peaks at just the west of HD\,164492\,A and have high $^{12}$CO $J$=2--1/$J$=1--0 line intensity ratios of $\sim$1.0, which corresponds to the kinetic temperature of $\sim$30\,K, indicating heating of the two clouds by HD\,164492\,A. 
The large velocity separation can neither be explained with gravitational binding by the total mass included in M\,20 nor the expanding motion driven by the feedback from HD\,164492\,A, and Paper I concluded that the two molecular clouds collided with each other $\sim0.3$\,Myr ago and triggered the formation of the central O star.

Although the association of the two clouds with M\,20 was firmly indicated by the increase of temperature toward the exciting O star, the spatial resolutions of the NANTEN and NANTEN2 dataset used in Paper I were much coarser than the optical and infrared images obtained toward M\,20, and detailed distributions and dynamics of the two colliding clouds have not been studied.
In this paper we present the results of our new observations of $^{12}$CO $J$=1--0 and $J$=3--2 at high angular resolutions of $22''$\,--\,$35''$.
The new dataset allows us to investigate the detailed evolutional scenario of M\,20 through the cloud-cloud collision.
Section 2 describes observations and Section 3 observational results of the two colliding clouds in M\,20. 
Section 4 makes a new analysis of the model data of the cloud-cloud collision in \citet{haw2015b, haw2015} and discusses comparisons with the observations.
Section 5 concludes the paper. The equinox of the celestial coordinates used in this paper is J2000.0. 

\section{Observations} \label{sec:obs}
\subsection{Mopra CO $J$=1--0 observations}  \label{sec:obs:1}
The Mopra 22-m telescope in Australia was used for the observations of the CO $J$=1--0 emission during 2011 October.
The backend system ``MOPS'' enabled us to obtain the three CO $J$=1--0 isotopes, $^{12}$CO $J$=1--0, $^{13}$CO $J$=1--0 and C$^{18}$O $J$=1--0 simultaneously, providing a velocity coverage and a velocity resolution of $360\kms$ and $0.08\kms$, respectively.
The OTF (on-the-fly) mode was used with a unit field of $4'\times4'$ toward a $8' \times 8'$ area of the M\,20 region.
The obtained spectra were smoothed along the velocity axis to a 0.44$\kms$ resolution.
The pointing accuracy was kept within 7$''$ by observing 86\,GHz SiO masers every 1 hour.
For the absolute intensity calibrations, Orion-KL (R.A., Dec.)=($-5^{\rm h}35^{\rm m}14\fs5$, $-5\degr22\arcmin29\farcs6$) was observed and compared with the CO spectra obtained by \citet{lad2005}, which was calibrated with an ``extended'' beam efficiency.
The typical r.m.s. noise fluctuations in the $^{12}$CO, $^{13}$CO and C$^{18}$O $J$=1--0 emission are 0.6\,K, 0.4\,K and 0.4\,K, respectively, with the typical system noise temperature of 400\,--\,600 K in the SSB. 

\subsection{ASTE CO $J$=3--2 observations}  \label{sec:obs:2}
Observations of the $^{12}$CO $J$=3--2 transition were performed with the ASTE 10-m telescope located in Chile in 2014 June \citep{eza2004, eza2008, ino2008}. 
The waveguide-type sideband-separating SIS mixer receiver ``CATS345'' and the digital spectrometer ``MAC'' were used at a frequency coverage of 128\,MHz and a frequency resolution of 0.125\,MHz resolution \citep{sor2000}, which corresponds to a velocity coverage and a velocity separation of 111$\kms$ and 0.11$\kms$, respectively, at 345\,GHz . 
The beamsize was 22$''$ at 345\,GHz, and the observations were made with the OTF mode toward the same area as the Mopra observations at a grid spacing of 7.5$''$.
The pointing accuracy was checked every $\sim1.5$\,hours to keep within 5$''$ by observing W\,Aql (R.A., Dec.)=($19^{\rm h}15^{\rm m}23\fs35$, $-7\degr02\arcmin50\farcs3$).
The absolute intensity calibration was made with observations of W28 (R.A., Dec.)=($18^{\rm h}00^{\rm m}30\fs4$, $-24\degr03\arcmin58\farcs5$). 
The day-to-day fluctuations of the peak intensity were within 10\,\%.
The typical system temperature was $\sim$250\,K in SSB and the final r.m.s noise fluctuations are typically 0.2\,K at the output velocity resolution 0.44$\kms$.

\section{Results}
\subsection{CO distributions}
\begin{figure}
\epsscale{.65}
\plotone{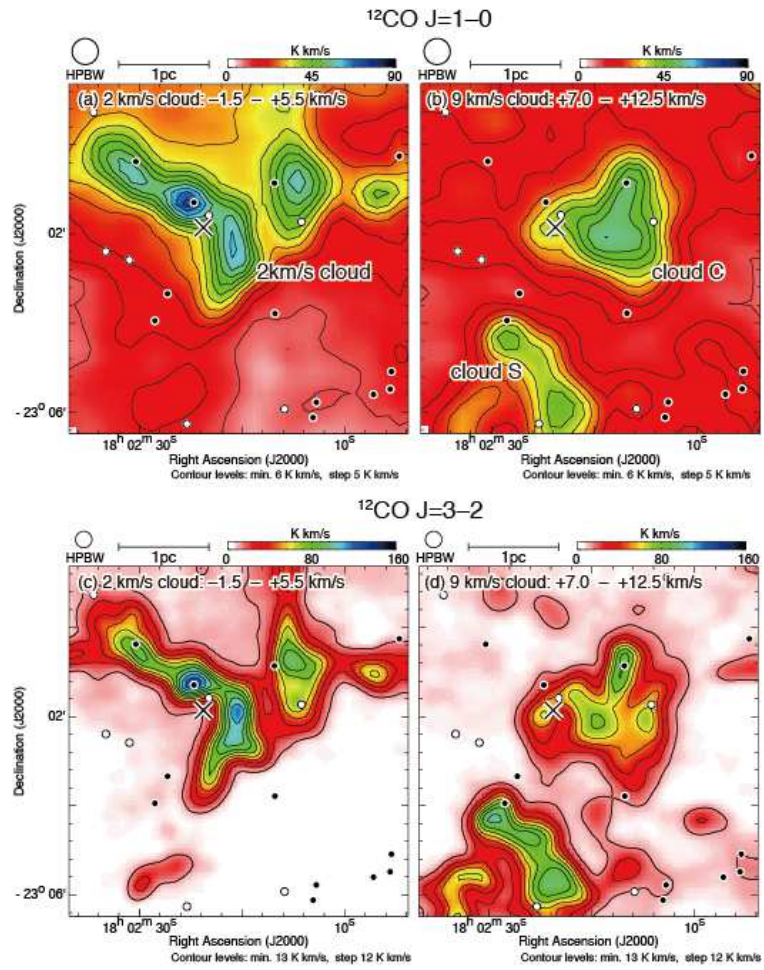}
\caption{Integrated intensity distributions of $^{12}$CO $J$=1--0 (upper panels) and $^{12}$CO $J$=3--2 (lower panels) for the two colliding clouds. Plotted symbols are the same as Figure\,\ref{opt}, but the Class 0/I stars are shown with filled black circles.
 \label{lb}}
\end{figure}

\begin{figure}
\epsscale{.9}
\plotone{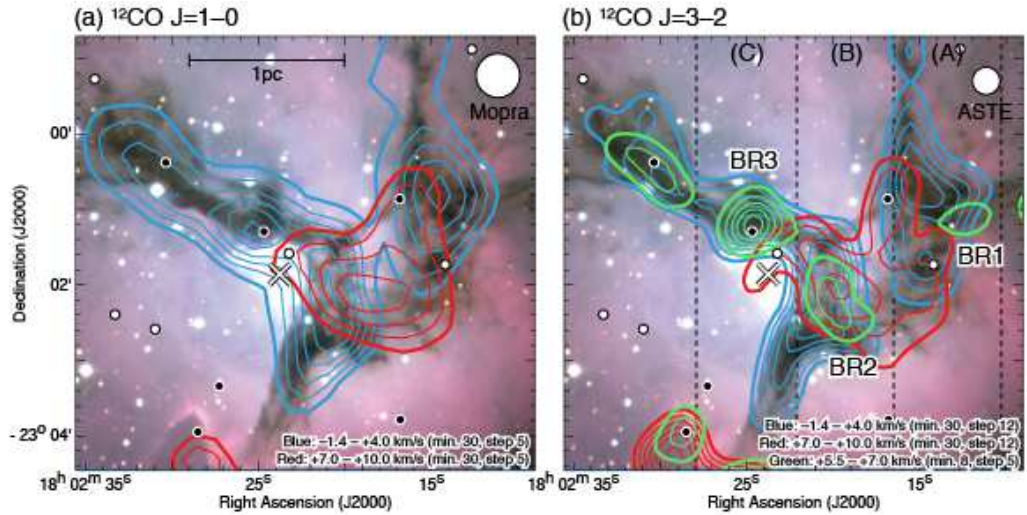}
\caption{The contour maps of the two colliding clouds are shown superimposed on the optical image of M\,20, where the $^{12}$CO $J$=1--0 emission is shown in (a)  and $^{12}$CO $J$=3--2 is in (b). $2\kms$ cloud and cloud C are plotted in blue contours and red contours, respectively. In (b) the bridge features BR1, BR2, and BR3 are added with green contours.  The velocity range and the contour levels are shown in the right-bottom of each panel. Dashed lines plotted in (b) indicate the integration ranges of the declination-velocity diagrams in Figure\,\ref{vb}.
 \label{lb_comp}}
\end{figure}

Figure\,\ref{lb} shows the CO $J$=1--0 and $J$=3--2 integrated intensity distributions of the two colliding clouds in M\,20.
The blueshifted cloud is pronounced at $-1$\,--\,5$\kms$, while the redshifted cloud is distributed at 7\,--12$\kms$. 
The two clouds are separated by $\sim7\kms$.
Following Paper I, we hereafter refer the blushifted cloud ``$2\kms$ cloud'' and the redshifted cloud ``9$\kms$ cloud''.
$2\kms$ cloud has an elongated distribution which streches along the east-west direction, and as presented later in Figure\,\ref{lb_comp}, it coincides with the dark lanes which lay in front of the M\,20 nebula.
Our new CO observations resolve the inner-structures of $2\kms$ cloud into several clumpy structures with sizes of 0.2\,--\,0.8\,pc. 
Compared with $^{12}$CO $J$=1--0, $^{12}$CO $J$=3--2 highlights these clumpy structures and is not sensitive to the diffuse CO emission widely distributed in the $^{12}$CO $J$=1--0 maps.

$9\kms$ cloud consists of several spatially separated clouds as discussed in Paper I, where two of them are included in the present observed region. 
 The cloud colliding with the $2\kms$ cloud is at the very west of HD\,164492\,A, (R.A., Dec.) $\sim$ ($18^{\rm h}2^{\rm m}18^{\rm s}$, $-23\degr2\arcmin$), which is referred to as ``cloud C'' in Paper I, while the other one ``cloud S'' is distributed at (R.A., Dec.) $\sim$ ($18^{\rm h}2^{\rm m}25^{\rm s}$, $-23\degr5\arcmin$).
Cloud S is not directly related to the collision (Paper I) and in this study we do not focus on this cloud, although it has been gaining attention because the optical jet HH\,399 is protruding from the protostellar core TC2 embedded around the northern tip of cloud S \citep{cer1998}.
While $2\kms$ cloud has overall similar distributions between $^{12}$CO $J$=1--0 and $^{12}$CO $J$=3--2, cloud C has different distributions; It has only one peak in the $^{12}$CO $J$=1--0 map (Figure\,\ref{lb}(c)) but has four local peaks in the $^{12}$CO $J$=3--2 map (Figure\,\ref{lb}(d)).

\begin{figure}
\epsscale{.8}
\plotone{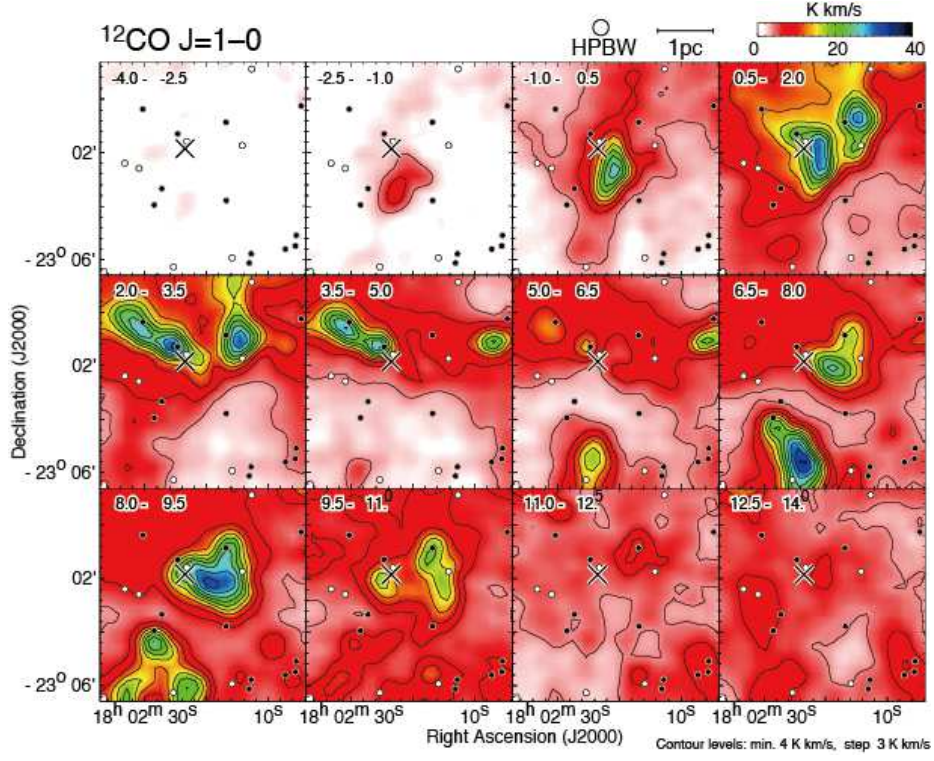}
\caption{Velocity channel map of the $^{12}$CO $J$=1--0 emission with a velocity step of $1.5\kms$. Plots are the same as Figure\,\ref{lb}.
 \label{channel12}}
\end{figure}

In order to investigate the physical interactions between $2\kms$ cloud and cloud C, in Figures\,\ref{channel12} and \ref{channel32} we present the $^{12}$CO $J$=1--0 and $^{12}$CO $J$=3--2 velocity channel maps over the entire velocity range of the two colliding clouds.
The $^{12}$CO $J$=1--0 emission in the intermediate velocity range of the two clouds (5\,--\,$6.5\kms$) is dominated by relatively weak, diffuse emission widely distributed above declination of $\sim-23\degr3\arcmin$, and small scale structures are hardly seen with significant detections.
The $^{12}$CO $J$=3--2, on the other hand, shows some clumpy components in the intermediate velocity range free from the diffuse emission.

\begin{figure}
\epsscale{.7}
\plotone{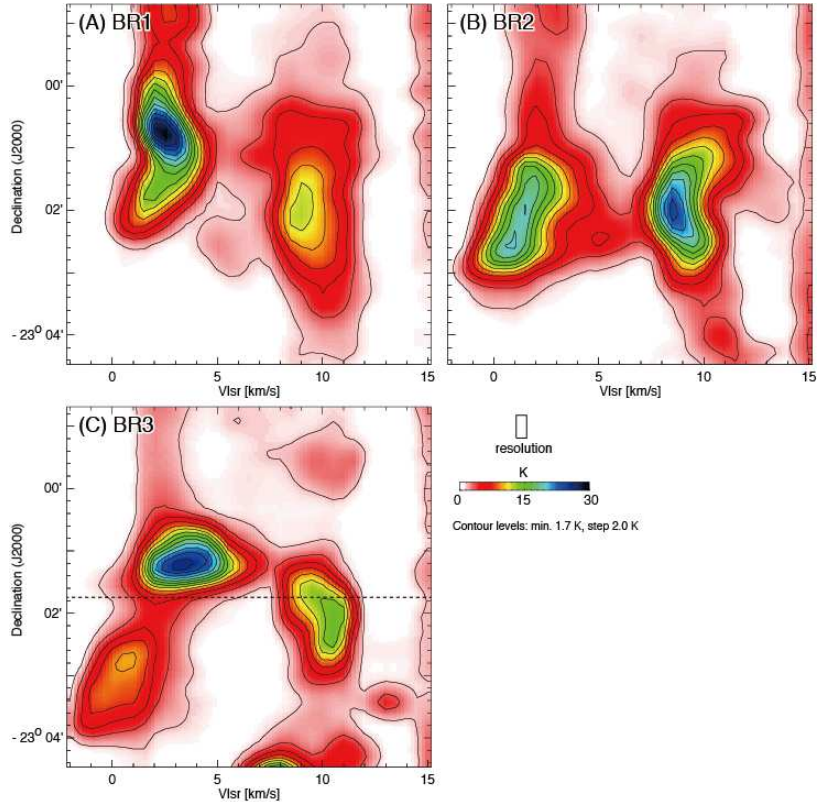}
\caption{Declination-velocity diagrams of the $^{12}$CO $J$=3--2 emission integrated over the ranges shown in Figure\,\ref{lb_comp}(b) with dashed lines. 
The dotted lines in (c) shows the declination of HD\,164492\,A.
 \label{vb}}
\end{figure}

Comparisons of the two colliding clouds and the intermediate velocity gas are shown in Figure\,\ref{lb_comp}, where the contour maps of $2\kms$ cloud (blue contours) and $9\kms$ cloud (red contours) are superimposed on the optical image, and the intermediate velocity features are shown only in the $^{12}$CO $J$=3--2 map (Figure\,\ref{lb_comp}(b)) in green contours.
The distribution of the $2\kms$ cloud is consistent with the dark lanes observable against the bright nebular background in the optical.
On the other hand, cloud C has no correspondence with the optical image and is likely located in the rear of the nebula.
It is interesting to note that $2\kms$ cloud and cloud C have complementary distribution; cloud C is sandwiched between the eastern component and the western component of $2\kms$ cloud.
The clumpy structures at the intermediate velocity range basically trace the gas distribution of $2\kms$ cloud.
We find that three of the clumpy structures, dubbed as BR1, BR2, and BR3 in Figure\,\ref{lb_comp}(b), can be identified as the bridge features which connect between $2\kms$ cloud and cloud C in velocity space.
This can be seen in the $^{12}$CO $J$=3--2 declination-velocity diagrams in Figure\,\ref{vb}.

\begin{figure}
\epsscale{.8}
\plotone{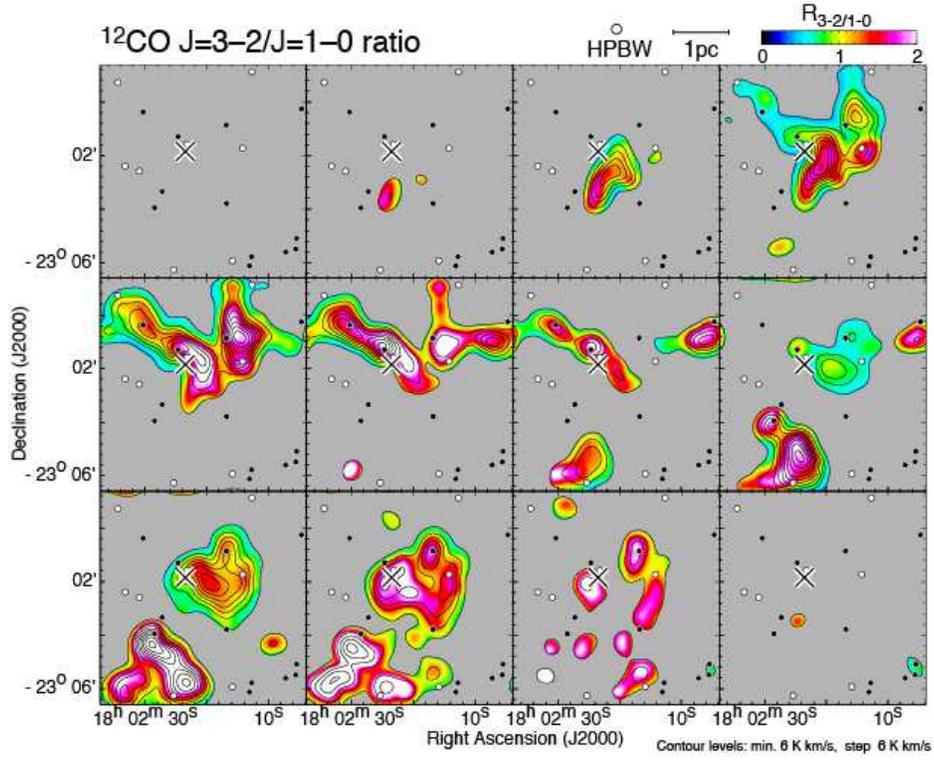}
\caption{Velocity channel map of the $^{12}$CO $J$=3--2/$J$=1--0 intensity ratio. Contours show the $^{12}$CO $J$=3--2 emission which was spatially smoothed to be 35$''$. Plotted symbols are the same as Figure\,\ref{lb}.
 \label{channel_ratio}}
\end{figure}

Figure\,\ref{channel_ratio} shows the velocity channel maps of the $^{12}$CO $J$=3--2/$J$=1--0 intensity ratio (hereafter $R_{3-2/1-0}$), where the color shows $R_{3-2/1-0}$ and contours indicate the $^{12}$CO $J$=3--2 distribution.
The two colliding clouds and the intermediate velocity gas have high $R_{3-2/1-0}$ of larger than 1.0, up to over 2.0, throughout the present target region.
That both the $2\kms$ and $9\kms$ cloud (cloud C and cloud S) are highly excited was already discussed in Paper I by taking ratios of $^{12}$CO $J$=2--1/$J$=1--0 ($R_{2-1/1-0}$) at a larger angular resolution of 4$'$.
The large velocity gradient analysis in Paper I indicates that these high $R_{2-1/1-0}$ gas is attributed to high kinetic temperature of the gas such as over 20\,K, which is significantly higher than the typical gas temperature of 10\,K in the galactic disk without heating, and following the discussion in Paper I, the high $R_{3-2/1-0}$ in clouds in M\,20 can also be interpreted as high gas temperature with $>20$\,K.
It is reasonable to deem that such high kinetic temperature is due to heating by HD\,164492\,A, illumination of the strong UV radiation and/or physical interaction with the ionized gas in the H{\sc ii} region.

\begin{figure}
\epsscale{.8}
\plotone{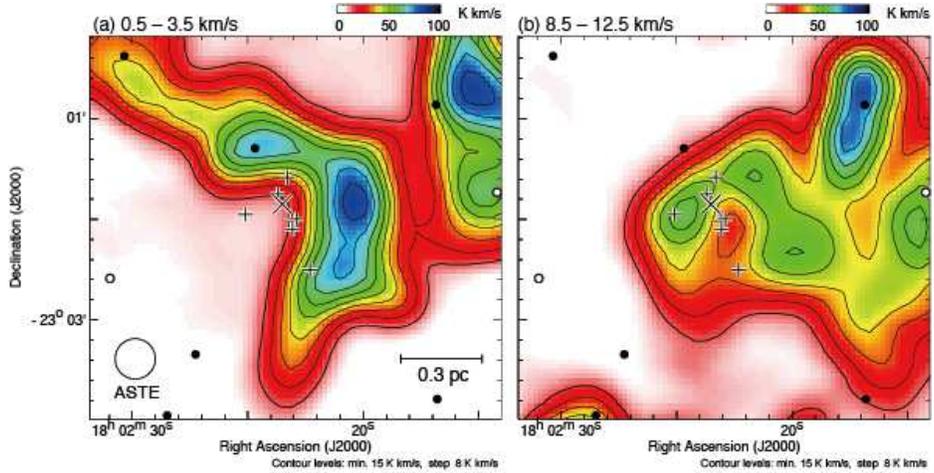}
\caption{The two hole structures surrounding the central stars of M\,20 are presented in $^{12}$CO $J$=3--2. Large cross indicates HD\,164492\,A, and black circles and white circles depict the Class I/0 and Class II YSOs \citep{rho2006}, respectively. Small crosses indicate the optical and infrared sources listed in Table\,\ref{tab:1}.
 \label{hole}}
\end{figure}

The disruptive feedback effects become more efficient at the neighborhood of HD\,164492\,A, and indeed the gas at the central a few pc of HD\,164492\,A has been completely dissipated as seen in Figure\,\ref{opt}(b), where the central stellar group including HD\,164492\,A is cleared from gas and dust, and is surrounded by the curved dark lane with a radius of 0.2\,--\,0.3\,pc. 
The northern part of the surrounding dark lane has a bright rim at its southern border, indicating the illumination by the strong UV radiation from HD\,164492\,A.
A molecular counterpart of the dark lane is seen in $2\kms$ cloud at 0.5\,--\,$3.5\kms$ as shown in Figure\,\ref{hole}(a).
In addition, we found another hole in cloud C at 8.5\,--\,$12.5\kms$ as seen in Figure\,\ref{hole}(b).
Different from the (incomplete) hole in $2\kms$ cloud, the hole in cloud C is closed with a radius of $\sim0.3$\,pc, consistent with that in the hole in $2\kms$ cloud.
Although its center does not coincide with HD\,164492\,A (it is shifted toward the south-west from HD\,164492\,A by $\sim0.1$\,pc), it is reasonable to deem that the second hole was also made by the feedback from HD\,164492\,A. 
This offers another support for the coexistence of $2\kms$ cloud and cloud C within the M\,20 nebula.

\subsection{Comparisons with infrared sources}
\begin{figure}
\epsscale{.8}
\plotone{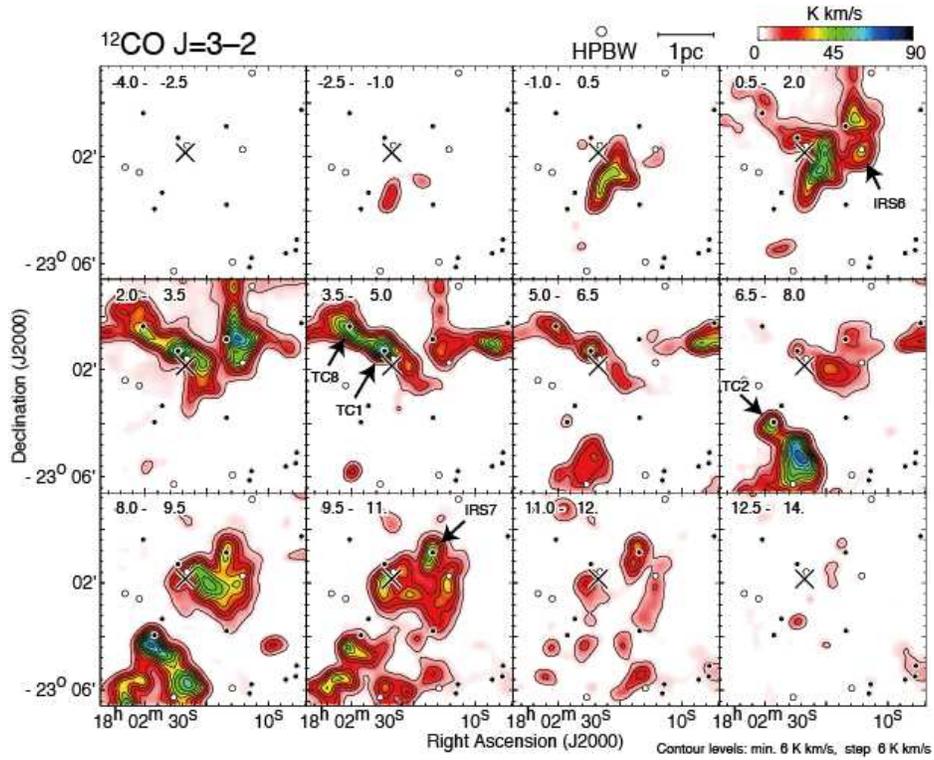}
\caption{Velocity channel map of the $^{12}$CO $J$=3--2 emission. Plots are the same as Figure\,\ref{lb}. The infrared sources (IRS6 and IRS7) and sub-mm sources (TC1, TC2, and TC8), which are likely associated with the $^{12}$CO $J$=3--2 peaks, are depicted by arrows.
 \label{channel32}}
\end{figure}

Associations of the infrared sources with the molecular clouds are investigated with the the velocity channel maps in Figures\,\ref{channel12}, \ref{channel32}, and \ref{channel_ratio}.
We found five infrared sources which show coincidence with the $^{12}$CO $J$=3--2 clumps (see arrows in Figure\,\ref{channel32}).
Three of them are the YSOs toward the dense dust cores TC1, TC2, and TC8, and associations of the class 0/I YSOs with these cores were discussed in \citet{lef2008}.
TC1 and TC8 are seen in $2\kms$ cloud, and TC2 is in cloud S in $9\kms$ cloud. Note that TC2 corresponds to the bridge feature BR3 as seen in Figure\,\ref{lb_comp}.
On the other hand, associations of the compact CO emission toward the remaining two sources, which corresponds to IRS6 and IRS7, were not found in the dust continuum observations in \citet{lef2008}.
These authors discussed that IRS6 and IRS7 are in latest stages of early stellar evolution when the parental envelope of the newly born stars has been almost fully photoevaporated, and hence the enhancement of the $^{12}$CO $J$=3--2 emission toward these two sources may be attributed to high temperature of the remaining parental gas.

\section{Discussion}

Our new CO data obtained with the ASTE and Mopra telescopes have revealed that:
\begin{enumerate}
\item From heating signatures, $2\kms$ cloud and cloud C are both located in the vicinity of M\,20. However they are both associated with distinct features of M\,20. $2\kms$ cloud which coincides with the dark lanes are in front of the nebula along the line-of-sight, while $9\kms$ cloud is in the rear.
\item $2\kms$ cloud and cloud C show complementary distributions.  Cloud C is located at the gap between the eastern component and the western component of $2\kms$ cloud. 
\item In the position-velocity diagrams, $2\kms$ cloud and cloud C are connected with each other by the bridge features. We identified three bridge features BR1, BR2, and BR3 with the $^{12}$CO $J$=3--2 data.
\end{enumerate}

\subsection{New analysis on the numerical calculations of the cloud-cloud collision}
\begin{figure}
\epsscale{1.}
\plotone{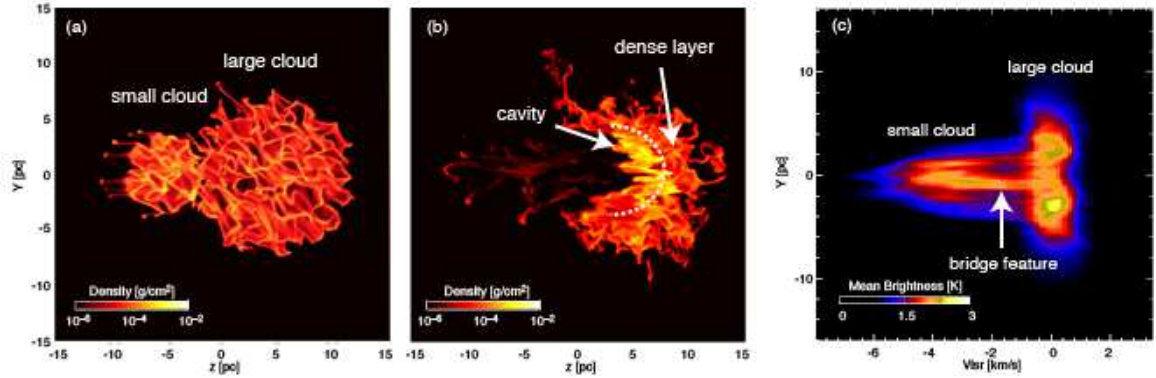}
\caption{Surface density plots of the cloud-cloud collision simulations by \citet{tak2014}. (a) the clouds prior to a $10\kms$ collision. (b) A snapshot of the collision, where the time corresponds to the maximum number of formed cores (see \citealt{tak2014} for details). Both of the two snapshots are at a viewing angle perpendicular to the collision axis. (c) The position-velocity diagram of the synthetic CO $J$=1--0 data of the 10$\kms$ collision model shown in (b). The viewing angle of the synthetic observation is set to be along the collision axis such that the clouds are coincident along the line-of-sight.
 \label{model}}
\end{figure}

In order to interpret these observational signatures, we here make comparisons with the theoretical works on cloud-cloud collision.
\citet{tak2014} calculated collisions between two Bonnor Ebert spheres of different size, seeded with turbulence.
The surface density plots of the collision model with a colliding velocity of $10\kms$ are shown in Figures\,\ref{model}(a) and \ref{model}(b).
The velocity separation of the two colliding clouds in M\,20 is measured as $\sim7\kms$.
Considering the viewing angle of the collision, $10\kms$ is a reasonable assumption of the colliding velocity in M\,20.
Figure\,\ref{model}(a) shows a snapshot prior to the collision with a viewing angle perpendicular to the collision axis, where the large cloud is at rest and the small cloud is moving rightward at $10\kms$ .
Figure\,\ref{model}(b) shows a snapshot at the time when the maximum number of self-gravitating cores are formed.
A cavity being created by the collision is seen in the large cloud.
At this time, the small cloud completely streamed into the dense shell at the interface of the collision, which corresponds to the stage 2 of Figure\,\ref{sch}.

\citet{haw2015} carried out synthetic CO observations with the Takahira et al.'s $10\kms$ collision model.
The authors postprocessed the data shown in Figure\,\ref{model}(b) with the {\tt\string TORUS} radiation transport and hydrodynamics code and produced synthetic $^{12}$CO $J$=1--0 cube data with $(x, y, v)$ axes (see \citealt{haw2015} for the detailed set-ups).
The output data has a spatial resolution of 0.2\,pc and a velocity resolution of $0.04\kms$.
Figure\,\ref{model}(c) shows a position-velocity diagram of the resulting $^{12}$CO $J$=1--0 data.
The viewing angle of the synthetic observations was set to be along the collision axis such that the clouds are coincident along the line-of-sight, and the position-velocity map was made by integrating along the entire $X$-axis.
The two clouds seen in the position-velocity map are separated by $\sim4\kms$, which indicates the deceleration of the colliding velocity, and the bridge features which connect the two clouds are seen at the intermediate velocity range ($-3$\,--\,$-1\kms$).

\begin{figure}
\epsscale{.5}
\plotone{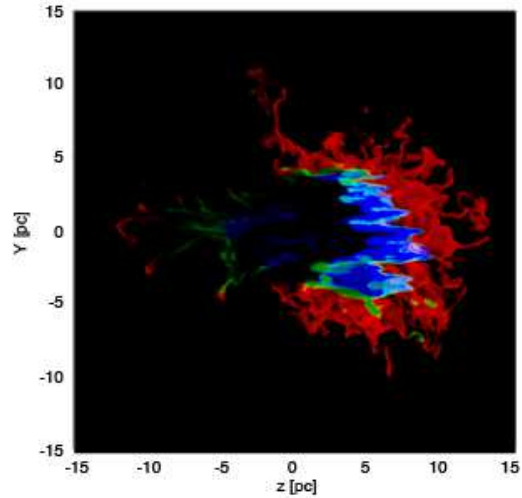}
\caption{Same as Figure\,\ref{model}(b) but the density components within different $v_{\rm z}$ ranges are shown in different colors. Red indicates $v_{\rm z} \geq 1\kms$, which corresponds to the large cloud. Blue shows the small cloud components with $v_{\rm z} < -3\kms$. Green is for the bridge feature at $-3\kms \leq v_{\rm z} < 1\kms$.
 \label{model_3col}}
\end{figure}

In Figure\,\ref{model_3col} we break down the surface density plot shown in Figure\,\ref{model} into three velocity components, i.e., the large cloud at $v_{\rm z} \geq 1\kms$ is shown in red, the small cloud at $v_{\rm z} < -3\kms$ is in blue, and the bridge features at $-3 \leq v_{\rm z} < 1$ in green.
It is seen that the bridge feature is the turbulent gas which is excited at the thin boundary between the small cloud ( = dense layer inside the cavity) and the large cloud.
\citet{ino2013} discussed that the turbulence as well as the magnetic field can increase the effective Jeans mass to orders of magnitude above the thermal Jeans mass, leading to mass accretion rates high enough to form massive stars.
Although the calculations by \citet{tak2014} did not include the magnetic fields, the thin turbulent layer at the interface of the two colliding cloud may provide sites of the high-mass star formation.

\begin{figure}
\epsscale{.7}
\plotone{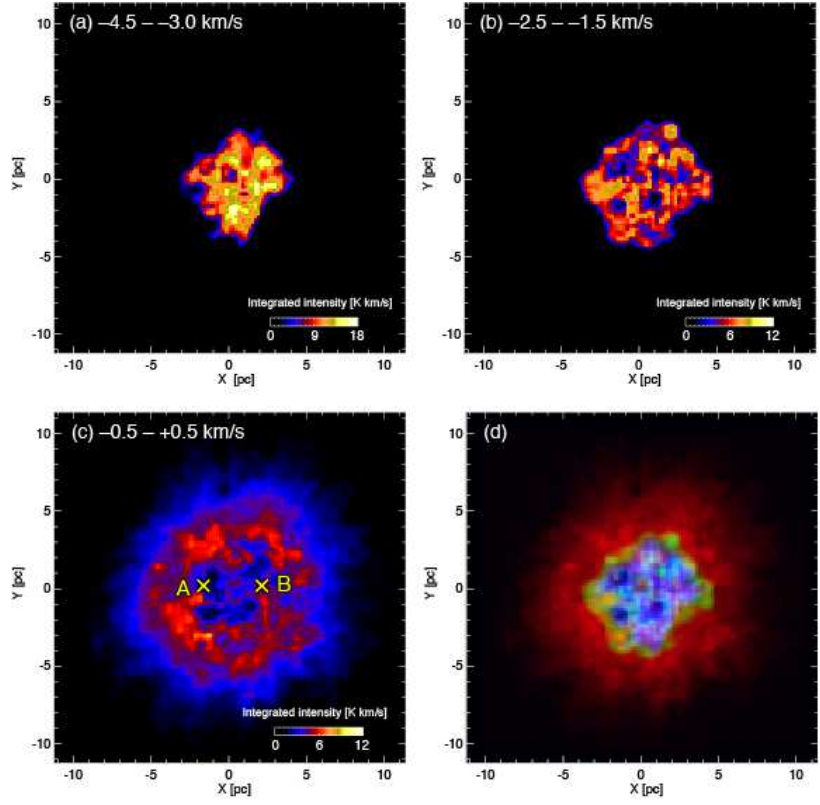}
\caption{Spatial distributions of the synthetic CO $J$=1--0 data of the collision model in Figure\,\ref{model}(b) are shown in (a)--(c) for three different velocity ranges. 
The velocity components of the small cloud and the large cloud are shown in (a) and (c), respectively, while the components at the intermediate velocity between the two clouds is in (b). The two crosses indicate the positions of point A and point B whose spectra are shown in Figure\,\ref{modelspec}.
(d) The three color composite image of the three velocity components in (a)--(c). red: (a), green: (b), blue: (c). 
 \label{model_th}}
\end{figure}

Figure\,\ref{model_th} shows the intensity distributions of the synthetic CO data on the $x$-$y$ plane integrated over the velocity ranges of the small cloud (a), the bridge feature (b), and the large cloud (c).
The large cloud overall has a ring-like distribution, which is due to the cavity created by the collision as seen in Figure\,\ref{model}(b). 
The small cloud which corresponds to the dense compressed layer which is caving the large cloud is compact and bright in the CO emission.
The bridge feature, or the thin turbulent layer at the interface of the collision, consists of several filamentary structures having widths ranging from 0.2\,pc to 1\,pc.
For comparisons, a three color composite image of the small cloud (blue), the bridge feature (green), and the large cloud (red) is presented in Figure\,\ref{model_th}(d).
Interestingly, the small cloud coincides with the inside of the ring of the large cloud, showing complementary distribution between the two colliding clouds.
The full extent of the bridge feature is slightly larger than that of the small cloud, and hence it looks being a thin layer which is surrounding the small cloud.

\begin{figure}
\epsscale{.7}
\plotone{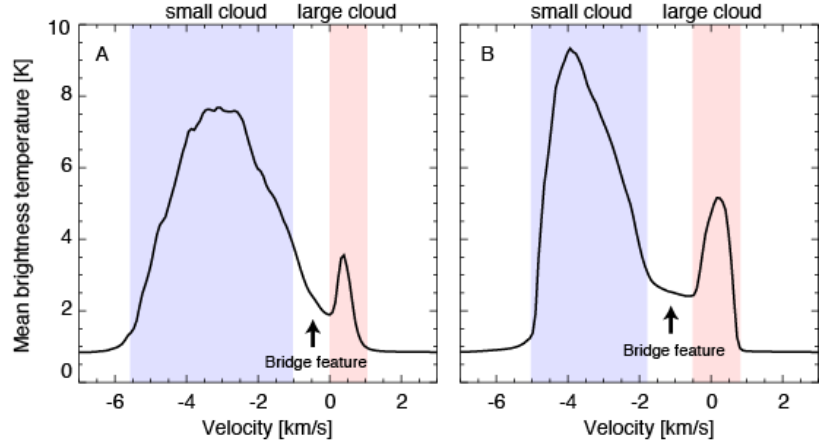}
\caption{Example spectra of the synthetic CO $J$=1--0 data toward the crosses in Figure\,\ref{model_th}(c). The velocity ranges of the small cloud and the large cloud are shaded by blue and red, respectively. The bridge features at the intermediate velocity range have flattened profiles and can be distinguished from the components of small cloud and the large cloud.
 \label{modelspec}}
\end{figure}

Figure\,\ref{modelspec} shows the typical CO spectra of the synthetic $^{12}$CO $J$=1--0 cube, where the two points depicted by crosses in Figure\,\ref{model_th}(c) are selected.
In the spectrum at point A, it is difficult to uniquely identify the bridge feature due to the contamination of the broadened emission from the small cloud.
At point B, on the other hand, the bridge feature can be distinguished as a flattened profile from the emission of the two colliding clouds.

\subsection{Comparisons between observations and model}

\begin{figure}
\epsscale{.95}
\plotone{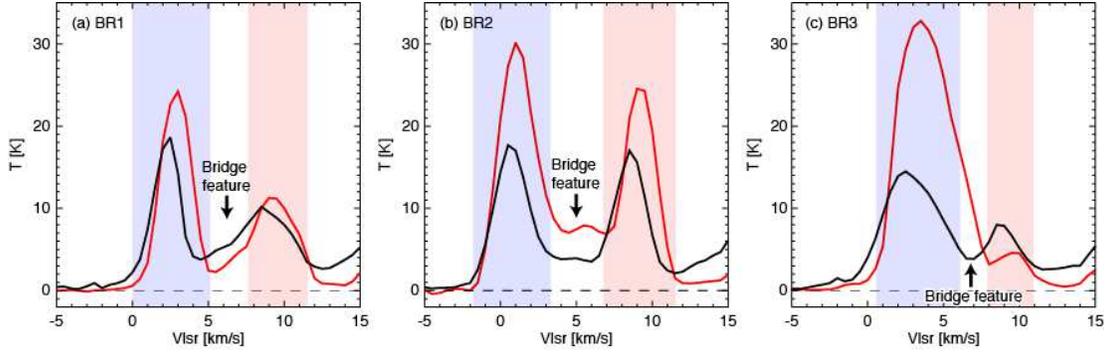}
\caption{$^{12}$CO $J$=1--0 (black) and $^{12}$CO $J$=3--2 (red) spectra toward the bridge features BR1, BR2, and BR3. The velocity ranges of $2\kms$ cloud and cloud C are shaded by blue and red, respectively.  \label{spec}}
\end{figure}

These characteristic features seen in the numerical calculations of cloud-cloud collision, i.e., the complementary distribution of the two colliding clouds, the bridge feature at the intermediate velocity range, and its flattened CO spectrum, can be seen in the present CO $J$=1--0 and $J$=3--2 observations.
The complementary distribution between the two colliding clouds is presented in Figure\,\ref{lb_comp}, where cloud C is sandwiched between the eastern component and the western component of $2\kms$ cloud.
The three bridge features BR1\,--\,BR3 identified in the $^{12}$CO $J$=3--2 data (Figure\,\ref{lb_comp}(b)) are distributed at the rim or just at the outside of cloud C.
The CO spectra of the three bridge features are shown in Figure\,\ref{spec}.
The profiles of the BR1 and BR2 spectra are similar to the model profile in the point B in Figure\,\ref{modelspec}, which has a flattened profile at the intermediate velocity range, while the profile in BR3 looks similar to that at point A.
All of these observed signatures lend strong support for the cloud-cloud collision scenario in M\,20.

We note that the numerical calculations by \citet{tak2014} do not include the magnetic fields.
\citet{ino2013} pointed out that the magnetic field plays a critical role to form the dense cores in the turbulent layer at the interface of the collision.
The future numerical calculations of the cloud-cloud collision including the magnetic field is necessary to fully understand the morphological structures and kinematics of the turbulent layer (= bridge features) between the colliding clouds.
In addition, the present spatial resolutions of the ASTE and Mopra data are not high enough to resolve the bridge features in M\,20 to investigate the filamentary inner-structures as seen in the model data (Figure\,\ref{model_th}(b)).
However, it is interesting to note that the dark dust lanes seen in the optical image in Figure\,\ref{opt} show entangled filamentary structures toward the complementary distribution between $2\kms$ cloud and cloud C.
Future ALMA observations will allow us to investigate the inner-structures of the colliding regions to make detailed comparisons with the theoretical works to reproduce the dense clumps, precursors of the O stars.

\subsection{Cloud-cloud collision in M\,20}

In Table\,\ref{tab} we summarize the physical parameters of the cloud-cloud collisions measured in the previous studies and in this study.
The masses of $2\kms$ cloud and cloud C in M\,20 are each estimated to be $10^3$\,$\msun$ for an assumed distance of 1.7\,kpc, respectively, where we assume an X-factor, the empirical conversion factor from the integrated intensity of $^{12}$CO $J$=1--0 to the H$_2$ column density, of $2\times10^{20}$\,cm$^{-2}$\,(K\,km\,s$^{-1}$)$^{-1}$ \citep{str1998}. These figures are consistent with the estimate in Paper I.
The H$_2$ column density is measured to be $\sim10^{22}$\,cm$^{-2}$ for both of the two clouds.
\citet{fuk2016a} discussed that H$_2$ column density is the critical parameter in the cloud-cloud collision scenario which determines the difference between the formation of single O star and massive star clusters, and $10^{23}$\,cm$^{-2}$ is the threshold to form the massive star clusters such as RCW\,38 via cloud-cloud collision.
Our results in M\,20, which harbors one single O star, are consistent with the discussion by \citet{fuk2016a}.
Compared with other regions listed in Table\,\ref{tab}, M\,20 is the case in which the masses and the column densities of the colliding clouds are smallest, although the classification of the formed O star is earlier than the O8V or O9V star in RCW\,120.

Timescale of the cloud-cloud collision in M\,20 can be calculated with the physical separation and the relative velocity of the two colliding clouds.
Since $2\kms$ cloud and cloud C are both embedded within the H{\sc ii} region of M\,20, the distance between the two clouds are limited to be within 3\,--\,4\,pc, which is the size of the nebula in M\,20.
If we tentatively assume a viewing angle of the collision in M\,20 as 45$^\circ$, the relative velocity of the two colliding clouds is $\sim10\kms$, then the timescale of the collision is estimated to be less than 0.3\,--\,0.4\,Myr, which is consistent with the age of M\,20 measured from the size of the H{\sc ii} region by \citet{cer1998}.

Formation of the low-mass stars is another intriguing topic in M\,20 \citep{lef2002,lef2008,rho2006}.
\citet{lef2008} discussed that the filaments which trisect the optical nebula in M\,20 were probably self-gravitating too before the birth of M\,20, and that the fragmentation of the filament which may lead to the YSO formation can be accounted for by MHD-driven instabilities, not by the shock interactions with the expanding H{\sc ii} region.
The growth timescale of the MHD instabilities is estimated to be as large as 1\,Myr, suggesting that the fragmentation would have started before the onset of the photoionization.
However, it may well be that the fragmentation occurred at an early evolutionary stage of the cloud-cloud collision. This hypothesis raises a possibility that the cloud-cloud collision which triggered the formation of HD\,164492\,A might accelerate the subsequent formation and evolution of the cores in the fragments.
It should be interesting to note a speculation that this is the case in the cold dust core TC1 in the eastern lane of the filaments, which harbors a class 0/I YSO \citep{lef2008}.
Our results indicate that the bridge feature BR\,3 is spatially well correlated with TC1 as shown in Figure\,\ref{lb_comp}.
The existence of the bridge feature indicates the existence of the turbulent gas excited by the cloud-cloud collision, suggesting a possibility that the TC2 core evolved under the condition affected by the cloud-cloud collision, although we cannot exclude that the central object of TC2 was already well evolved before the collision.

\section{Summary}

The conclusions of the present study are summarized as follows;
\begin{enumerate}
\item We performed high resolution CO $J$=1--0 and CO $J$=3--2 observations toward the two colliding molecular clouds in the galactic H{\sc ii} region M\,20 with Mopra and ASTE. The two clouds are resolved into details, allowing us to make direct comparisons with the optical image of M\,20. 
\item We identified two peculiar molecular gas structures in the two colliding clouds, $2\kms$ cloud and cloud C. One is the spatially complementary distribution. Cloud C coincides with the gap at the center of $2\kms$ cloud and appears to being sandwiched by the eastern component and the western component of $2\kms$ cloud. The other is the bridge feature which connects the two clouds in velocity space. We identified three bridge features BR1\,--\,BR3. They are located around the sides of cloud C. These structures strongly indicates that, although they are separated by $\sim7\kms$, the two colliding clouds are physically associated with each other.
\item In order to interpret these observed structures, we make a new analysis on the model CO $J$=1--0 data generated by \citet{haw2015} with the $10\kms$ collision model calculated in \citet{tak2014}. As a result, we reproduce the similar gas structures found in the present observations such as the complementary distributions and the bridge features, where the complementary distribution can be accounted for by the cavity created by the collision between two dissimilar clouds and the bridge feature as the turbulent gas excited in the thin layer at the interface of the collision. Our new results lend support for the cloud-cloud collision scenario in M\,20.
\end{enumerate}

This work was financially supported by Grants-in-Aid for Scientific Research (KAKENHI) of the Japanese society for the Promotion of Science (JSPS; grant numbers 15H05694, 15K17607, 24224005, 26247026, 25287035, and 23540277). We also acknowledge the support of the Mitsubishi Foundation and the Sumitomo Foundation. TJH is funded by an Imperial College London Junior Research Fellowship


\clearpage

\begin{table*}
\begin{center}
\caption{Identification of components in the HD 164492 Complex \citep{rho2008}. } \label{tab:1}
\begin{tabular}{lcll}
\hline
\hline
Object & Position (J2000) & Type & Comments \\
(1) & (2) & (3)  & (4)\\
\hline
A			& $18:02:23.5, -23:01:51$		& O7.5V(III)			& the exciting star \\
B			& $18:02:23.7, -23:01:45$		& A2Ia				&  \\
C (IRS\,1)		& $18:02:23.1, -23:02:01$		& B6V				& hard X-rays \\
D (IRS\, 2)	& $18:02:22.9, -23:02:00$		& Be, LkH$\alpha$\,123	& proplyd \\
E			& $18:02:23.1, -23:02:06$		& F3V				&  \\
F			& $18:02:25.1, -23:01:57$		& 					& optical source \\
G (IRS\,4)		& $18:02:22.3, -23:02:30$		&  					& optical source \\
IRS\,3 (HST\,1)	& $18:02:23.3, -23:01:35$		& late-F to mid-G		& proplyd \\
IRS\,5		& $18:02:21.1, -23:01:04$		&					& IR source \\
IRS\,6		& $18:02:14.1, -23:01:44$		&					& IR source, proplyd \\
IRS\,7		& $18:02:16.8, -23:00:52$		&					& IR source, proplyd \\

\hline
\end{tabular}
\end{center}
\end{table*}

\begin{table}
\footnotesize
\begin{center}
\caption{Comparisons between the cloud-cloud collision regions.}
\label{tab}
\begin{tabular}{ccccccccc}
\tableline\tableline
 \multirow{2}{*}{Name}	& \multirow{2}{*}{cloud mass}	 &   molecular		&   relative & complementary	&  bridge & \multirow{2}{*}{cluster age} & \multirow{2}{*}{\# of O stars	} &  \multirow{2}{*}{reference} \\
 					&				 & column density	& velocity	 & distribution		& feature & \\
 					&  [$\times10^3\,M_\odot$]		 &  [$\times10^{22}$\,cm$^{-2}$]		&   [km\,s$^{-1}$] 	&&\\
 (1)	&  (2)  & (3)  &  (4)  &  (5)  &  (6)  & (7) & (8) & (9)\\
 \tableline
RCW\,38    & (20, 3)  & (10, 1) & 12 & no & yes & $\sim$0.1& $\sim20$& [1]  \\
NGC\,3603 & (70, 10) & (10, 10) & 15 & no & yes & $\sim$2& $\sim30$ & [2] \\
Westerlund\,2 & (90, 80) & (20, 2) & 16 &yes & yes & $\sim$2 & 14 & [3, 4]\\
$[$DBS2003$]$179 & (200, 200) & (8, 5)  & 20 & yes & yes & $\sim$5 & $>10$ & [5] \\
 ONC (M\,42)	 & (20, 3) & (20, 1) & \multirow{2}{*}{$\sim$7$^{\rm (a)}$} & yes & no & \multirow{2}{*}{$<1$ }& $\sim$10 & \multirow{2}{*}{[6]} \\
 ONC (M\,43)	 & (0.3, 0.2) & (6, 2) & & yes & no &  & 1 & \\

 RCW\,120 & (50, 4) & (3, 0.8) & 20 & yes & yes & $\sim$0.2 & 1 & [7] \\
 N159W-South & (9, 6) & (10, 10) & $\sim$8$^{\rm (b)}$ & no & no & $\sim$0.06 & 1 & [8]\\
 N159E-Papillon & (5, 7, 8) & (4, 4, 6) & $\sim$9$^{\rm (c)}$ & no & no & $\sim$0.2 & 1 & [9]\\
 M\,20 & (1, 1) & (1, 1) & 7.5 &yes & yes & $\sim0.3$& 1 & This study, [10]\\

\tableline
\tableline
\end{tabular}
\tablecomments{Column: (1) Name. (2, 3) Molecular masses and column densities of the two/three colliding clouds. (4) Relative velocity between the colliding clouds. (7, 8) Age and the number of O stars. (9) References: [1] \citet{fuk2016a} [2] \citet{fuk2014}, [3] \citet{fur2009}, [4] \citet{oha2010}, [5] Kuwahara et al. (2016) in prep. [6] Fukui et al. (2016b) in prep. [7] \citet{tor2015}, [8] \citet{fuk2015b}. [9] \citet{sai2016} [10] \citet{tor2011}. (a)--(c) corrected for the projection.}
\end{center}
\end{table}



\end{document}